\documentclass[%
  onecolumn
]{mpi2015-cscpreprint}

\usepackage[american]{babel}

\usepackage{graphicx}
\usepackage[compatibility=false]{caption}
\usepackage{subcaption}
\usepackage{tikz}
\usetikzlibrary{calc, positioning, hobby, fit, arrows.meta}
\definecolor{mardiorange}{HTML}{D0662B}
\definecolor{mardiblue}{HTML}{005EAA}
\definecolor{background}{HTML}{EEEEEE}

\usepackage{booktabs, makecell} 
\usepackage{enumitem}
\usepackage{float, caption}
\usepackage{hyperref}
\usepackage{listings} 
	\lstdefinelanguage{json}{
		basicstyle=\normalfont\ttfamily,,
		showstringspaces=false,
		breaklines=false,
		frame=lines,
		backgroundcolor=\color{background},
	}
\usepackage{url}
\usepackage[color=violet!80, textcolor=white]{todonotes}

\usepackage{amssymb}
\usepackage{amsthm}

\newcounter{mymac@matlab}
\newcommand{\matlab}{MATLAB%
  \ifnum\value{mymac@matlab}<1%
  \textsuperscript{\textregistered}%
  \setcounter{mymac@matlab}{1}%
  \fi%
}
\newcommand{\octave}{\textsf{GNU Octave}}

\begin{document}
  

\title{Towards a Benchmark Framework for Model Order Reduction in the Mathematical Research Data Initiative (MaRDI)}
  
\author[1]{Peter Benner}
\affil[1]{Max Planck Institute for Dynamics of Complex Technical Systems, Sandtorstr.~1, 39106 Magdeburg\authorcr
  \email{\{benner, lund, saak\}@mpi-magdeburg.mpg.de}; \orcid{0000-0003-3362-4103}, \orcid{0000-0001-9851-6061}, \orcid{0000-0001-5567-9637}}
  
\author[1]{Kathryn Lund}
  
\author[1]{Jens Saak}
  
\shorttitle{MaRDIMark}
\shortauthor{P.~Benner, K.~Lund, J.~Saak}
\shortdate{}
  

  
\abstract{The race for the most efficient, accurate, and universal algorithm in scientific computing drives innovation. At the same time, this healthy competition is only beneficial if the research output is actually comparable to prior results. Fairly comparing algorithms can be a complex endeavor, as the implementation, configuration, compute environment, and test problems need to be well-defined. Due to the increase in computer-based experiments, new infrastructure for facilitating the exchange and comparison of new algorithms is also needed. To this end, we propose a benchmark framework, as a set of generic specifications for comparing implementations of algorithms using test cases native to a community. Its value lies in its ability to fairly compare and validate existing methods for new applications, as well as compare newly developed methods with existing ones.
As a prototype for a more general framework, we have begun building a benchmark tool for the model order reduction (MOR) community.

The data basis of the tool is the collection of the Model Order Reduction Wiki (MORWiki). The wiki features three main categories: benchmarks, methods, and software. An editorial board curates submissions and patrols edited entries. Data sets for linear and parametric-linear models are already well represented in the existing collection. Data sets for non-linear or procedural models, for which only evaluation data, or codes / algorithmic descriptions, rather than equations, are available, are being added and extended. Properties and interesting characteristics used for benchmark selection and later assessments are recorded in the model metadata.

Our tool, the Model Order Reduction Benchmarker (MORB) is under active development for linear time-invariant systems and solvers.  An ontology (MORBO) and knowledge graph are being developed in parallel. They catalog benchmark problem sets and their metadata and will also be integrated into the Mathematical Research Data Initiative (MaRDI) Portal\footnote{\url{https://www.mardi4nfdi.de/}}, to help improve the findability of such data sets. MORB faces a number of technical and field-specific challenges, and we hope to recruit community input and feedback while presenting some initial results.}

\novelty{We present a preliminary software tool for benchmarking model order reduction software on linear, time-invariant, first-order systems.}

\maketitle
\section{Introduction}%
\label{sec:intro}
The Mathematical Research Data Initiative (MaRDI)\footnote{\url{www.mardi4nfdi.de}} is a consortium of the National Research Data Initiative (NFDI)\footnote{\url{www.nfdi.de}}, whose overarching goal is to improve and promote responsible research data management practices in the German scientific landscape and beyond.  MaRDI itself concentrates on several fields of mathematics featured as content-specific task areas (TA):
\begin{enumerate}[label=({TA}\arabic*), leftmargin=5\parindent] \itemsep0em
	\item Computer Algebra,\hfill{\small (exact data)}\label{ta1}
	\item Scientific Computing,\hfill{\small (inexact data)}\label{ta2}
	\item Statistics and Machine Learning, and\hfill{\small (uncertain data)}\label{ta3}
	\item Cooperation with Other Disciplinces.\hfill{\small (mixed data and interoperability)}\label{ta4}
\end{enumerate}
Within TA2, we focus primarily on numerical algorithms and their software implementations, which are used to compute approximate solutions and simulations of scientific models.  Via collaboration between the Max Planck Institute for Dynamics of Complex Technical Systems and the University of M\"unster, TA2 addresses the following measures (M):
\begin{enumerate}[label=({M}\arabic*), leftmargin=5\parindent] \itemsep0em
	\item Knowledge graph of numerical algorithms;\label{m1}
	\item Open interfaces for scientific computing;\label{m2}
	\item Benchmark framework; and\label{m3}
	\item Description and design of Findable, Accessible, Interoperable, and Reproducible (FAIR)\footnote{\url{www.go-fair.org/fair-principles}} workflows for computational science and engineering (CSE).\label{m4}
\end{enumerate}
For this report, we will concentrate on progress in~\ref{m3}.

\section{MaRDIMark}%
\label{sec:mardimark}
We begin with a domain-independent specification of a benchmark workflow, MaRDIMark.  MaRDIMark is broken into five modules, as displayed in Figure~\ref{fig:mardimark}.  The content and function of each module is detailed in the following sections.

\newlength{\mygap}
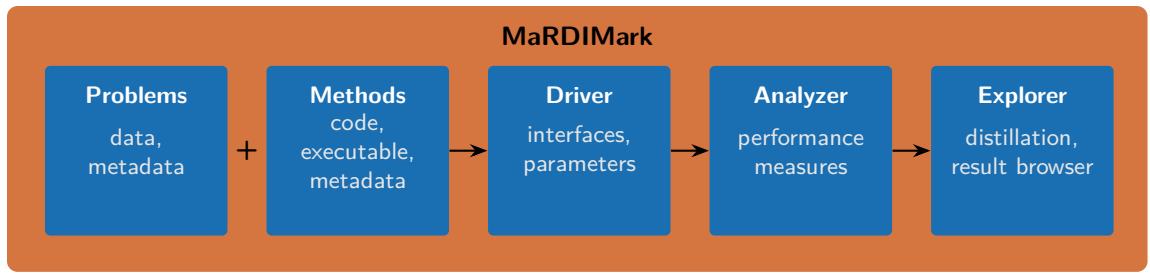
\begin{figure}[H]
	\setlength{\mygap}{0.0333\linewidth}%
	{\fontfamily{lmss}
		\begin{tikzpicture}[%
			toplevel/.style={%
				rectangle,
				fill=mardiorange!90,
				minimum width=\linewidth,
				minimum height=.235\linewidth,
				rounded corners
			},
			bottomlevel/.style={%
				rectangle,
				minimum width=.475\linewidth,
				minimum height=.24\linewidth,
				rounded corners,
			},
			element/.style={%
				rectangle,
				fill=mardiblue!90,
				minimum width=.16\linewidth,
				minimum height=.15\linewidth,
				rounded corners=2pt,
			},
			label/.style={below=1ex, text=white, font=\small\bfseries},
			layer/.style={above=.1ex, text=white, font=\footnotesize\bfseries}
			]
			\node[toplevel] (main) {};
			\node[below=1ex] at (main.north) {\bfseries MaRDIMark};
			\node[element, right=\mygap of main.west, yshift=-1ex] (problems) {%
				\begin{minipage}{.13\linewidth}\small\color{black!10}\centering
					data,
					metadata
				\end{minipage}
			};
			\node[label] at (problems.north) {Problems};
			%
			\node[element, right=\mygap of problems] (methods) {%
				\begin{minipage}{.13\linewidth}\small\color{black!10}\centering
					code,
					executable,
					metadata
				\end{minipage}
			};
			\node[label] at (methods.north) {Methods};
			%
			\node[element, right=\mygap of methods] (driver) {%
				\begin{minipage}{.13\linewidth}\small\color{black!10}\centering
					interfaces,
					parameters
				\end{minipage}
			};
			\node[label] at (driver.north) {Driver};
			%
			\node[element, right=\mygap of driver] (analyzer) {%
				\begin{minipage}{.13\linewidth}\small\color{black!10}\centering
					performance measures
				\end{minipage}
			};
			\node[label] at (analyzer.north) {Analyzer};
			%
			\node[element, right=\mygap of analyzer] (explorer) {%
				\begin{minipage}{.13\linewidth}\small\color{black!10}\centering
					distillation,
					result browser
				\end{minipage}
			};
			\node[label] at (explorer.north) {Explorer};
			%
			\path (problems.east) -- (methods.west) node[midway]{\textbf{+}};
			\draw[-Stealth, thick] (methods.east) -- (driver.west);
			\draw[-Stealth, thick] (driver.east) -- (analyzer.west);
			\draw[-Stealth, thick] (analyzer.east) -- (explorer.west);
		\end{tikzpicture}
	}
	\caption{Overview of MaRDIMark.}\label{fig:mardimark}
\end{figure}

\subsection{Problems}\label{sec:problems}
This module consists of the benchmark problems themselves, both as mathematical abstractions and as datasets stored on a computer.  Note that datasets might take the form of both assembly scripts and static floating-point numbers.  A good benchmark collection should minimally comprise a database of unique problem IDs, download links, mathematical descriptions, fields referring to domain-specific mathematical and numerical properties, file formats, license data, creator and editor identifiers (ideally ORCID\footnote{\url{www.orcid.org}}), and any other information that helps with data provenance, findability, and accessibility.  Ontologies and knowledge graphs like AlgoData\footnote{A deliverable of~\ref{m1}. See \url{https://algodata.mardi4nfdi.de/}.} can add value to such a database, as they allow users to query complex relations, instead of just look up values in a table.  For example, a user may want to know which methods can be applied to a given benchmark, which brings us to the next module.

\subsection{Methods}%
\label{sec:methods}
The Methods module is similar to the Problems module, except that the objects of interest are instead mathematical procedures and algorithms for solving a class of problems.  The definition of a method for a particular field can be nontrivial, especially when all implementation details are taken into account. For example, most methods in numerical mathematics are initially presented as pseudocode, which a practitioner must translate into actual code to use on a computer.  Minimally, the practitioner must choose hardware (i.e., a physical machine), an operating system (and version), and a programming language (and version).  Within that language, there are multiple ways to execute any mathematical operation, and each may have slightly different floating-point behavior, which further depends on what libraries the language and operating system are running on the backend.  We loosely refer to a method, its pseudocode, and its implementation choices as an \emph{algorithm isotope}, in analogy to elemental isotopes.  Each community must decide on its own how fine-grained their algorithm isotopes should be, and these definitions will necessarily vary from application to application, depending on what aspects of a class of methods are being studied.

\subsection{Driver}%
\label{sec:driver}
The linchpin of a benchmark workflow is a routine (or set of routines) that executes specified methods on specified problems, i.e., the Driver module.  In practice, a benchmark workflow may be written in a general-purpose language but need to call routines written in another language, without introducing unnecessary overhead.  Here open interfaces developed by~\ref{m2} can be applied, which facilitate communication between different software packages and programming languages.

\subsection{Analyzer}%
\label{sec:analyzer}
Benchmarking by definition is concerned with the comparison of methods with respect to different measures, such as, time to solution, accuracy, quality, optimized cost function, and so forth.  We refer to the collection and calculation of these measures as the Analysis module.  A community must determine on its own which measures are relevant and how to implement them in an unbiased manner, so that no method is artificially penalized by the procedure used to calculate a measure.

\subsection{Explorer}%
\label{sec:explorer}
Finally, it is important that practitioners are able to visualize the results and share them with colleagues.  The Explorer module thus comprises graphs, plots, tables, and reports that MaRDIMark synthesizes.  Convenient formats for many mathematical fields might be encoded in TeX, HTML, or Markdown, or as a Jupyter notebook.

\section{Model Order Reduction Benchmarker (MORB)}%
\label{sec:morb}
To demonstrate MaRDIMark in action, we have begun developing a benchmark tool for the field of MOR, relying on the curated benchmark database hosted in the MORWiki~\cite{morWiki}.

\subsection{MORWiki}%
\label{sec:morwiki} 
MORWiki was launched in 2012 as a platform for the MOR community. It hosts established benchmark collections like the Oberwolfach Collection~\cite{morOWF} and SLICOT~\cite{morVar99b, morVar01} and new datasets. Currently, approximately 100 distinct datasets have been documented for a variety of linear, nonlinear, and parametric real-world and test examples.  Various methods have also been documented, along with a comparison table for MOR software and a BibTeX file for MOR literature.  Since the start of MaRDI, a searchable database for problem metadata has been compiled, along with an initial ontology that will be integrated into AlgoData and the MaRDIPortal\footnote{\url{https://portal.mardi4nfdi.de/}}.  Eventually, a search tool will be published that allows users to look up benchmarks with desired properties.

A number of challenging tasks remain open for MORWiki.  In particular, licenses are missing for older datasets like SLICOT and the Oberwolfach, despite the fact that they have been treated as if they were part of the public domain for nearly 20 years.  This poses a unique legal situation, especially if we upload updated versions of the data, to the MORWIKI Zenodo community\footnote{\url{https://zenodo.org/communities/morwiki/}} that conform to the file format standards of MORB (cf.\ Section~\ref{sec:morb_design}). Furthermore, it is important that such a database is built by the larger community, even if the overall editorial work is maintained by a small group.  Incentivizing researchers to not only publish their data, but also prepare it in a specific format to improve interoperability is difficult. A key selling point is that papers with public code and data accumulate higher citations over time~\cite{ColHSetal20}.  Finally, much work remains regarding the mathematical metadata for MORWiki benchmarks.  Systems theory enjoys a plethora of problem varieties, which must often be determined by hand or via numerical eigensolvers, which can be prohibitively expensive for large systems.

\subsection{Design specifications}%
\label{sec:morb_design}
For the initial prototype of MORB, we have aimed to keep things as simple as possible.  We have focused on linear, time-invariant, first-order systems (LTI-FOS) of the form
\begin{align*}
	E\dot x(t) &= A x(t) + B u(t),\\
	y(t) &= C x(t) + D u(t),
\end{align*}
where $A, E \in \mathbb{R}^{n \times n}$, $B \in \mathbb{R}^{n \times m}$, $C \in \mathbb{R}^{q \times n}$, and $D \in \mathbb{R}^{q \times m}$.  When $E$ and $C$ are not explicitly specified, the identity matrix is assumed for both; when $D$ is not specified, it is assumed to be zero.  For interoperability, we require that all the matrices for a given benchmark are stored as \texttt{.mat} files, version 7, which is supported by \matlab{}\footnote{\url{https://de.mathworks.com/help/matlab/import_export/mat-file-versions.html}} and SciPy\footnote{\url{https://docs.scipy.org/doc/scipy/reference/generated/scipy.io.loadmat.html}}, as well as \octave\footnote{\url{https://octave.org/}}, Julia\footnote{\url{https://julialang.org}}, and can be linked into most compiled languages via libmatio\footnote{\url{https://github.com/telehan/libmatio}}.  We have ensured that all LTI-FOS benchmarks in MORWiki are converted to such a format, and will be made public\footnotemark[7], as soon as the open licensing questions have been resolved.

A key component of MORB is the configuration file.  It is written in JSON\footnote{\url{https://www.json.org/}}, a format compatible with most widely used programming languages, and is relatively straightforward to set up and modify. The highest level must correspond to the filename (i.e., benchmark ID). The secondary level should minimally have the \texttt{"alg\_iso"} field and optional fields
\begin{itemize} \itemsep0em
	\item \texttt{"meas\_opt"} (options for Analyzer module);
	\item \texttt{"bode\_opt"} (options for Bode plots; part of Explorer module);
	\item \texttt{"plot\_opt"} (general options for plots; part of Explorer module); and
	\item \texttt{"report\_opt"} (options for final report; part of Explorer module).
\end{itemize}

See Figure~\ref{fig:config_file} for an example configuration file.  In this example, the solver \texttt{bt} (Balanced Truncation) is being configured for the packages \texttt{cst} (Control Systems Toolbox)\footnote{\url{www.mathworks.com/products/control.html}}, \texttt{emgr} (Empirical Grammian Framework)~\cite{morHim22}, \texttt{mess} (Matrix Equation Sparse Solver)~\cite{morBenKS21}, \texttt{morlab} (Model Order Reduction Laboratory)~\cite{morBenW21b}, and \texttt{pymor} (Model Order Reduction with Python)~\cite{morMilRS16}.  Depending on the package, different parameters like \texttt{tol} (Hankel singular value tolerance) and \texttt{max\_order} (maximum order of the reduced model) may be set.  Note that actually two algorithm isotopes are specified for both the \texttt{cst} and \texttt{mess} packages, because two different \texttt{tol} options have been provided.  Because \texttt{emgr} is set to \texttt{null}, six total algorithm isotopes will be run for \texttt{bt}.

\begin{figure}
	\input{newEngland_config}
	\caption{Example of a configuration file for the benchmark problem
      \texttt{newEngland\_n66m1q1}.}%
      \label{fig:config_file}
\end{figure}

For this version of MORB, the encasing workflow has been written in \matlab{}\@. This choice has allowed us to minimize interface issues, as most MOR software for this class of models is written in MATLAB\footnote{\url{https://morwiki.mpi-magdeburg.mpg.de/morwiki/index.php/Comparison_of_Software}}.  Furthermore, the MATLAB package MORLAB contains well documented routines for computing standard errors in the $L_0$, $L_1$, $L_2$, $L_{\infty}$ and $\mathcal{H}^2$ norms, as well as Bode, Frobenius, and Sigma plots, which we have relied on for the Analyzer module.

\subsection{Early prototype and sample reports}%
\begin{figure}[H]
	\centering
	\resizebox{.65\textwidth}{!}{\includegraphics{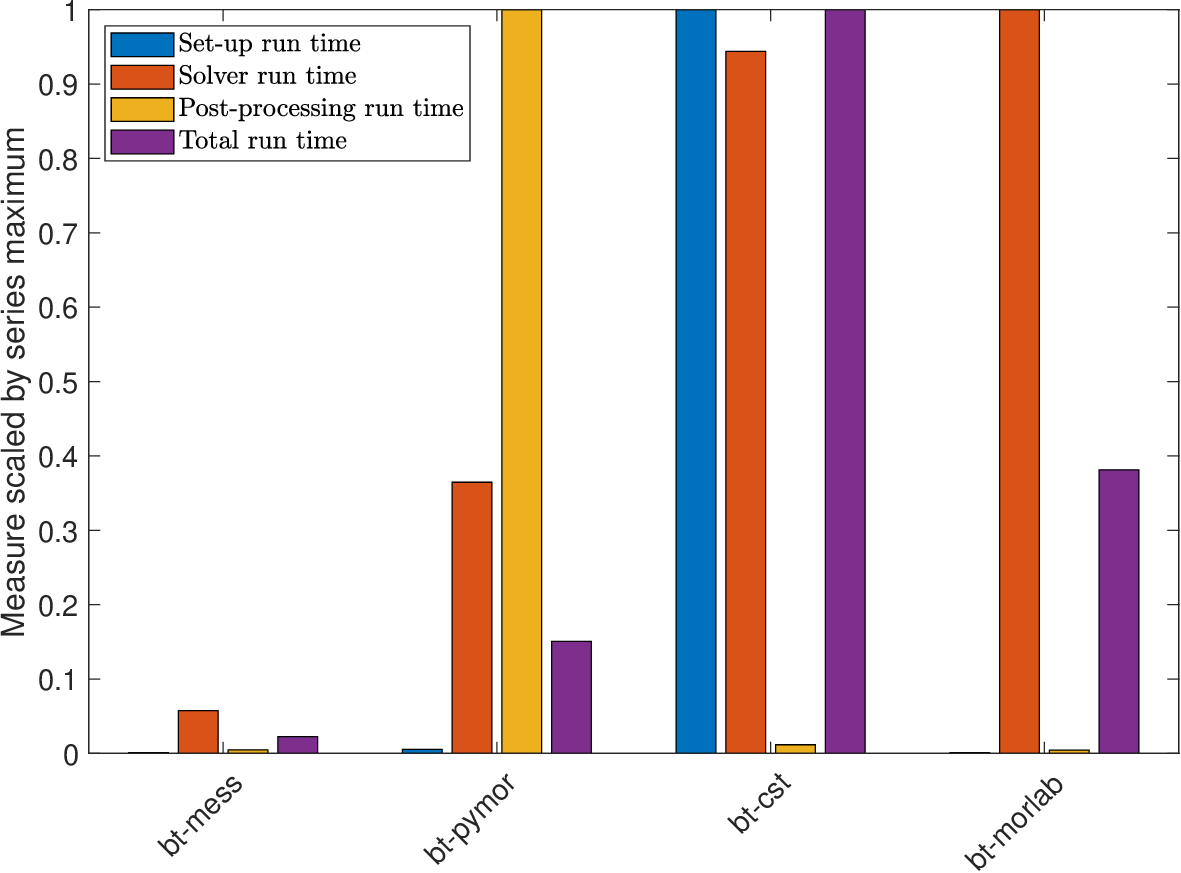}}
	\caption{Sample run time bar chart for
      \texttt{steelProfile\_n1357m7q6}. Note that timings are scaled by the
      maximum in the series.%
      \label{fig:steelProfile_runtime}}
\end{figure}

\label{sec:morb_sample}
A prototype of MORB that works on the steel profile benchmark~\cite{SaaB21} can be found on Zenodo~\cite{LunSB23}.  Furthermore, sample reports generated from other datasets are included there that demonstrate what the Explorer module looks like.  Details about the run itself-- a timestamp,  the benchmark ID, operating system, software version, and problem size-- are reported, along with a table and graph of the run times and a table of error norms, as well as error, Bode, Sigma, and Frobenius plots.  See Figure~\ref{fig:steelProfile_runtime} and Table~\ref{tab:steelProfile_error} for a preview of the full report.  In this version of MORB, reports are auto-generated as TeX files, which can then be compiled as PDFs.  Additional details about the run are stored in the comments of the TeX file, generated by the MATLAB command \texttt{system('systeminfo')}\footnote{This command only works on Windows. \url{https://www.mathworks.com/help/matlab/ref/system.html}}.

\begin{table}[H]
	\begin{center}
		\begin{tabular}{c|c|c|c}
			\toprule
			\thead{Algorithm Isotope} & \thead{$L_2$} & \thead{$L_{\infty}$} & \thead{$H_2$} \\
			\midrule
			bt-mess     & 2.23e-06       & 2.83e-06       & 4.66e-06       \\\hline
			bt-pymor    & 1.56e-05       & 1.78e-05       & 2.92e-05       \\\hline
			bt-cst      & 2.45e-04       & 1.95e-04       & 2.90e-03       \\\hline
			bt-morlab   & 2.86e-06       & 3.00e-06       & 5.55e-06       
		\end{tabular}
	\end{center}
	\caption{Sample error table for \texttt{steelProfile\_n1357m7q6}.%
    \label{tab:steelProfile_error}}
\end{table}

\section{Conclusions and Outlook}
We have demonstrated the versatility of MaRDIMark on a curated database of MOR benchmarks.  The resulting tool, MORB, is an initial attempt to compare solvers written in different languages with different syntax and parameter-parsing standards for a large class of benchmark examples.  There is still much room for improvement.  First of all, we plan to refactor it in Python, a free, open-source language with a large, active user-base.  This will ensure we comply with FAIR principles, especially accessibility and reproducibility, which the MATLAB version currently violates due to the requirement for paid licenses.  Secondly, the configuration file as described in Section~\ref{sec:morb_design} will be updated to remove redundancies in the hierarchy, to reflect better the modular design of the tool, and to enable parallelization for large numbers of algorithm isotopes.  In addition, we plan to integrate more algorithms from the field to allow for benchmarks on a wider variety of problems, particularly nonlinear and parametric examples.  Finally, the report itself will be updated to allow for a wider variety of formats (e.g., HTML and Markdown) and to incorporate progress from~\ref{m4} and MaRDIFlow~\cite{VelHB23}

Principles from MaRDIMark can already be seen in other benchmark tools.  One of the co-authors has actively incorporated the modular design in the stability analysis of communication-avoiding Gram-Schmidt and Krylov subspace methods~\cite{Lun22, LunCO23}.  Within MaRDI, TA3 has developed \texttt{mlr3}, which provides a unified interface for machine learning in the R language~\cite{BisSKetal23}.  The toolkit includes a benchmark feature\footnote{\url{https://mlr3book.mlr-org.com/chapters/chapter3/evaluation_and_benchmarking.html}}, which generates a grid of learner configurations, runs them on a set of tasks, and returns performance results that can be explored as a table or box plot.  Another well-developed open-source tool is \texttt{FitBenchmarking}~\cite{FitBench22}, which compares fitting and minimizer frameworks from software that is callable from Python.  The primary output is an interactive table, allowing users to quickly navigate results for interesting solver and benchmark combinations.  A key feature of both \texttt{mlr3} and \texttt{FitBenchmarking} is that they have been developed long-term by large teams of domain experts and software engineers.  Furthermore, both rely on curated benchmark databases established and maintained by their respective academic communities, and they both provide pathways for users to contribute new software to be benchmarked.

\section*{Acknowledgments}
We would like to thank Christian Himpe and Tim Mitchell for conversations that laid the foundation of the benchmark tool, as well as Alexander Stage and Malte Speidel, for work on the MORB ontology and search tool.  We further thank Martin K\"ohler for helping debug MORB.


\addcontentsline{toc}{section}{References}
\bibliography{mor, benchmarking}
\bibliographystyle{abbrvurl}
  
\end{document}